  \documentclass[preprint2]{aastex}

\slugcomment{draft 2000 Feb 15}

\shorttitle{Comparing Tycho--2 astrometry with UCAC1}
\shortauthors{Zacharias et al.}

\begin{document}

\title{Comparing Tycho--2 astrometry with UCAC1}

\author{N. Zacharias, M.I. Zacharias\altaffilmark{1} and S.E. Urban}
\affil{U.S.~Naval Observatory, 3450 Mass.~Ave.~NW, Washington, DC 20392}
\email{nz@pisces.usno.navy.mil}

\author{E. H{\o}g}
\affil{Copenhagen University Observatory, Juliane Maries Vej 30, DK-2100 Copenhagen, Denmark}

\altaffiltext{1}{also with Universities Space Research Association}


\begin{abstract}

The Tycho--2 Catalogue, released in
February 2000, is based on the ESA Hipparcos space mission data and 
various ground--based catalogs for proper motions.
An external comparison of the Tycho--2 astrometry is presented here
using the first U.S.~Naval Observatory CCD Astrograph Catalog (UCAC1).
The UCAC1 data were obtained from observations performed at CTIO
between February 1998 and November 1999, using the 206 mm aperture
5--element lens astrograph and a 4k x 4k CCD.
Only small systematic differences in position between Tycho--2 and UCAC1
up to 15 milliarcseconds (mas) are found, mainly as a function of
magnitude.
The standard deviations of the distributions of the position 
differences are in the 35 to 140 mas range, depending on magnitude.
The observed scatter in the position differences is about
30\% larger than expected from the combined formal, internal errors,
also depending on magnitude. 
The Tycho--2 Catalogue has the more precise positions for bright
stars (V$\le 10^{m}$) while the UCAC1 positions are significantly better
at the faint end ($11^{m} \le V \le 12.5^{m}$) of the magnitude range in common.
UCAC1 goes much fainter (to R$\approx 16^{m}$) than Tycho--2; however
complete sky coverage is not expected before mid 2003. 

\end{abstract}




\section{Introduction}

Two new major astrometric catalogs became available in early 2000.
The Tycho--2 Catalogue for the brightest 2.5 million stars
\citep{tycho2s}, \citep{tycho2l} and the first
U.S.~Naval Observatory CCD Astrograph Catalog, UCAC1, for
27 million stars on the Southern Hemisphere \citep{ucac1}.
Both catalogs are important steps towards the extension 
of the optical reference frame \citep{iau_jd} beyond densities
and magnitudes of the Hipparcos Catalogue.

The Tycho--2 Catalogue is a new, extended version of the original
Tycho Catalogue \citep{hip_tycho}, based on a re--reduction of
the ESA Hipparcos \mbox{space} mission Tycho data and 
over 140 ground--based catalogs for the Tycho--2 proper motions.
An external comparison of the Tycho--2 {\em astrometry} is presented here
utilizing the UCAC1 positions.
A similar comparison between the Tycho--1 Catalogue and CCD 
astrograph test data was \mbox{presented} earlier \citep{veniceT}.

The U.S.~Naval Observatory CCD Astrograph (UCAC) project was
planned and initiated in the mid 1990s \citep{gaus1}, \citep{1k_res},
\citep{veniceU}.
Observations started at the South Celestial Pole in 1998 
and full sky coverage is expected by mid 2003.
For the comparison presented in this paper, these new, high precision 
observations were available for about 80\% of the Southern Hemisphere,
covering the magnitude range R$\approx 8-16^{m}$.
Thus, particularly, the new, faint extension of the Tycho--2 catalog 
($11^{m}$ to $12.5^{m}$) is very well covered by these independent 
ground--based observations. 

Both catalogs are on the Hipparcos system, thus the International
Celestial Reference System (ICRS).
The epoch difference of about 8 years is bridged by proper motions 
given in the Tycho--2 catalog at the expense of introducing a third 
error contribution besides the positional errors of both catalogs.
The UCAC is not a photometric catalog, with only approximate
magnitudes given in a single bandpass (red).  Therefore {\em no} external
{\em photometric} comparison of the Tycho--2 data can be presented here.
Both Tycho--2 and UCAC1 are of great importance to the general
astronomical community and the astrometric comparison presented here 
is also of benefit for the Tycho--2 and in particular the UCAC projects.
Another important catalog comparison between the Tycho--2, the \mbox{ACT}
and the Hipparcos Catalogue is in preparation \cite{t2acthip}.

\section{Description of the Catalogs}

Here we briefly describe properties of both catalogs which are
relevant for this comparison.  For more details, the reader is
referred to the papers mentioned in the introduction.

\subsection{Tycho--2}

The major improvement of the Tycho--2 Catalogue over the original Tycho
Catalogue is the faint extension, providing positions for some 1.5 million
more stars, extending its limiting magnitude to about V=12.5.
This was made possible by a new reduction procedure which also
provides new, slightly improved positions for the stars previously
given in Tycho--1.
The mean epoch for the Tycho observations is in the range of
1990.72 to 1992.36, depending on the individual star.
The estimated precision of the Tycho positions is a function of
magnitude, as given in Table 1, adopted from \citep{tycho2s}.
An internal error of a position component for each star in Tycho--2
is derived from the distribution of the $\approx 130$ individual
observations per star made by the Hipparcos satellite.
Excluding the wings of the distribution, a standard error for the
position of each star is calculated.
The medians of those standard errors per magnitude bin are given 
in Table 1 ($\sigma_{T}$), providing a robust estimate of the standard 
error with outliers rejected.

Proper motions have been derived utilizing the Tycho--2 positions and
a large number of ground--based catalogs, including a new reduction
of the Astrographic Catalogue (AC) data.
Significant changes were made here, particularly in the magnitude dependent
systematic error corrections of the AC data as compared to the previous
release, the AC2000 \citep{ac2000}.
Assuming a mean difference of 7.7 years between the Tycho--2 and \mbox{UCAC1}
epochs and taking the formal errors of the Tycho--2 proper motions
from \citep{tycho2s}, we obtain the error contribution due to proper
motions, $\sigma_{PM}$, also given in Table 1.

\subsection{UCAC1}

The UCAC astrometric sky survey started in February 1998 at Cerro Tololo,
Chile.  
In this first, preliminary catalog, data are included up to November 1999,
thus the mean epoch of the catalog is about 1999.0.
The UCAC survey is performed in
a single bandpass (579-642 nm) with a 2-fold overlap pattern of
fields, with a long and a short exposure on each field.  
The 5--element lens astrograph of 206 mm aperture uses a single 4k by 4k 
CCD detector, providing a 61$\arcmin$ field of view with a
scale of 0.9"/pixel.
Stars in the range of 10 to 14th magnitude are the most precise
with an estimated positional error of $\approx$20 mas. 
Brighter stars are overexposed on the long exposures, hence only
the short exposures generally contribute to positions of stars
around magnitude 8 to 9.5.
\mbox{Stars} brighter than $8.0^{m}$ are also saturated on most short exposures.
For faint \mbox{stars}, the precision degrades 
to about 70 mas at the limiting magnitude of R$\approx$16.

A first catalog (UCAC1) for 27 million stars \citep{ucac1} was released
by the end of March 2000, see also \url{http://ad.usno.navy.mil/}. 
Preliminary proper motions are included for all stars in \mbox{UCAC1}.
Those proper motions have not been used here at all.
For the magnitude range overlapping with the Tycho--2 data,
the UCAC1 proper motions are mainly derived from AC and Tycho--2 data,
thus are highly correlated with the Tycho--2 proper motions.
For this catalog comparison the Tycho--2 proper motions were
used to update the Tycho--2 positions to the UCAC epoch of
each individual star in common.

The UCAC1 was reduced using Hipparcos, Tycho--1 and ACT data, thus
mainly depends on the original Tycho--1 Catalogue stars plus the
proper motions derived from the AC2000 reductions.
In particular, all new, faint stars in Tycho--2 which are 
not in Tycho--1 are field stars in UCAC1, thus were not used
as reference stars.
Therefore UCAC1 positions of those stars are largely independent of
Tycho--2 positions. 
Both Tycho--2 and UCAC1 are in the same system (Hipparcos),
except for a possible, slight alteration by the use of ACT 
proper motions.

All CCD frames of the UCAC1 data were reduced individually
using the astrometrically suitable reference stars on each frame
with a linear 6--parameter mapping model.
The low charge--transfer efficiency (CTE) of our CCD results
in significant coma--like systematic errors along the x--coordinate
(right ascension).
The effect in the raw data is up to $\pm$70 mas.
These errors have been corrected to first order.
Remaining systematic errors are believed to be on the 10 mas level.
The remaining residuals versus coma are $\le$ 10 mas for most part
in both axis.  There is an asymmetric feature of up to 20 mas in the
x--coordinate for a small fraction of the stars.  
These plots and more details can be found in the UCAC1 paper \citep{ucac1}.

A magnitude equation was already noticed in the residuals of the
CCD frame reductions, mainly for the x--coordinate.
However, for this first catalog neither corrections for a magnitude
term were applied nor were such terms included in the reduction
model. 
At this point it is not obvious if the magnitude dependent systematic
errors are inherent in the UCAC $x,y$--data or introduced by the 
ACT via proper motions.
Future UCAC astrometric reductions will use the Tycho--2 Catalogue,
which was not available at the time of UCAC1 reductions.

Weighted mean catalog positions were obtained from the individual
positions.  Precisions of the mean positions are
obtained from the scatter of the few ($\approx 2...6$) individual
observations per \mbox{star}.
The (squared) mean of these individual sigmas over \mbox{all} stars
in a magnitude bin is an estimate of the standard error of
the positional precision of the catalog and is given
in Table 1 as $\sigma_{U}$.
This $\sigma_{U}$ is slightly underestimated due to the correlations
of the individual field star positions from the short and long
exposure frames, which use a significant number of reference stars
in common for their reductions.

The precision of the UCAC1 positions is also a function of
magnitude.
For magnitudes $R_{U}\le$11, the precision of UCAC1 positions
is underestimated due to the correlation with the reference stars.
For the \mbox{UCAC} data the $R_{U}$ red magnitude has to be used.
On average, the color index is $V_{T} - R_{U} \approx 0.5$,
thus the precisions in position given in Table 1 for both
catalogs are for about the same stars on each line.
The last column in Table 1 gives the expected combined
standard error of (Tycho--2 $-$ UCAC1) position differences
as derived from internal, formal errors only.

\section{Results of the Catalog Comparison}

In total, close to 1.0 million stars have been found in common between 
the Tycho--2 Catalogue and UCAC1, adopting a match radius of 500 mas. 
Excluding the Tycho--1 stars (reference stars of \mbox{UCAC1}) and adopting
the same match radius we find 597,809 stars in common.

\subsection{Systematic errors}

Position differences (Tycho--2 $-$ UCAC1), at the UCAC1 epoch,
are plotted versus Tycho V magnitude in Figure 1.
All stars in common are shown here.
Using only the stars not included in Tycho--1 gives just the
faint part of the plots shown in Figure 1. 
One dot represents the mean of 2000 stars.
A magnitude equation is clearly present, with systematic
differences up to $\pm 15$ mas over the range of $\approx$5 magnitudes.
While the magnitude equation for the right ascension component (x)
is almost linear, it is non--linear for the declination 
component (y).

The systematic errors as a function of magnitude vary with
declination as can be seen in Figure 2.
Here the (Tycho--2 $-$ UCAC1) position
\mbox{differences} are plotted versus declination.
The general offset is due to the fact that the majority of
stars (faint end) have a non--zero position difference
due to the magnitude equation. 
The variation of this offset and thus the magnitude equa\-tion
follows a zonal pattern in declination.
A plot of the differences vs.~right ascension shows no pattern.
There are also no additional significant systematic errors of
the position differences with respect to color.
Figure 3 shows plots for the $\Delta \alpha \cos \delta$
and $\Delta \delta$ components vs.~$(B-V)$ Tycho color index.

\subsection{Error budget}

In the following statistics, only the \mbox{non-Tycho-1} stars are used.
Table 2 summarizes the results by magnitude bins, with all
position items given in units of mas.
The listed $V_{T}$ magnitude is for the center of each bin,
while $n$ gives the number of stars available per magnitude bin.
The standard deviations given in Table 2 are obtained from

\[  \sigma_{x} \ = \ \sqrt{\frac{\sum (x_{i} - \bar{x})^{2}}{n} }  \]

and similarly for the $y$--coordinate, where $x_{i}$,
$y_{i}$ are the individual (Tycho--2 $-$ UCAC1)
position differences and $\bar{x}, \bar{y}$ are
their arithmetic means for that magnitude bin.
In order to exclude outliers, two approaches were made.
First the $(x_{i} - \bar{x})^{2}$ values were sorted and
the largest 10\% were rejected.  
The resulting standard errors from this cut of the distribution
are given in Table 2 in the $\sigma_{cx}, \sigma_{cy}$ columns.
The second method used the 25 to 75 percentile of the distribution
of the position differences around their mean.
Assuming the central parts of our position differences
distributions to be Gaussian, $\sigma_{qx}, \sigma_{qy}$ 
values were derived (see Table 2).
For easy comparison, the last column of Table 2 gives the {\em expected}
combined standard error for our position differences, from the formal
errors alone, which has been copied from Table 1.

\section{Discussion}

\subsection{Magnitude equation}

A comparison with the Yale Southern Proper Motion (SPM) data \citep{spm}
shows that the apparently linear magnitude equation in \mbox{UCAC1} for 
the right ascension component does not extend to fainter magnitudes.
Results are presented in \citep{ucac1},
indicating good agree\-ment between UCAC1 and SPM positions around 15th
magnitude for both coordinates.
The systematic pattern as a {\em function of declination}
can best be understood to originate from the proper motion
part in this comparison.  Individual zones in the AC might
be responsible for this signature.
A plausible magnitude dependent error of $\approx$100 mas at a 1900
AC epoch would lead to an error in the ACT proper motions consistend
with a 10 mas offset seen in our comparison at the 1999 UCAC1 epoch.

The overall offset in Figure 2 results from an average slope in the
magnitude equation and the fact that the mean magnitude of the stars
in comparison here is fainter than the mean magnitude of the reference
stars used in the UCAC1 reductions.
This {\em mean} magnitude equation at the 8 to 12 mag range is possible
to come from the UCAC $x,y$--data. 
Many images of those stars are close to the pixel saturation level,
within a factor of $\approx$3 in the amplitude counts in either the
long or short exposure.
It is also conceivable, that remaining systematic errors associated
with the CTE effect show up as pure magnitude terms.
A systematic error in the Tycho--2 data of $\approx$ 5 to 10 mas
at the faint end can not be excluded from this comparison either.
However this is unlikely based on studies made in the Tycho--2 
construction (comparison with Hipparcos). 
Future reductions of the UCAC data will help to clarify this issue.

\subsection{Error budget}

There are 5 contributions to the observed scatter in the
position differences:
random errors in both the Tycho--2 and UCAC1 positions at their mean epochs, 
random errors in the proper motions needed to bridge the epoch difference
between the catalogs, additional noise from incompatibilities
and systematic differences.
The additional noise from incompatibilities includes items like
unknown multiplicity of stars, which shifts the center of light as a
function of time and bandpass.
It is conceivable that a relatively small number of multiple stars 
can bias the statistics significantly, particularly in the highest
precision area around 10th magnitude.
Unresolved components on the few 100 mas level are likely to
cause offsets in the photocenters on the 10 to 100 mas level, introducing 
a significant additional error.
Neither the cut of the observed distribution ($\sigma_{cx}, \sigma_{cy}$) 
nor the 50\% quantile standard errors are likely to eliminate
those problem stars from the statistics.
Also, positional shifts by differential refraction are not taken
into account in the UCAC data.  Individual stars, depending on their
color and zenith distance at the time of observation, will have
different offsets w.r.t.~the mean of the reference stars,
contrary to the Tycho--2 data which are collected outside the
Earth's atmosphere.
The difference in position between an O5 and an M5 main sequence star
due to differenctial color refrection is about 10 mas for the UCAC
bandpass and a zenit distance of 45$^{\circ}$.
Thus there are physical reasons to expect a noise term in the observed
scatter of the position differences in addition to the 3 internal 
errors (catalog 1, catalog 2, proper motions), even in the
absence of significant systematic errors.

For magnitudes V$\le$9.0 the statistics are affected by
the small number of available stars.
Overall the observed scatter in the position differences \mbox{agrees}
very well with the expected, formal, errors at those magnitudes.
In general the x--coordinate (right ascension) shows a significantly
larger error than the y--coordinate.
This very likely shows still unresolved systematic errors
in the UCAC data due to the CTE problems along the x--axis.

The largest discrepancies (typical 40 to 50\% underestimation of
errors) between observed and 
expected standard errors are seen in the mid range of 
fainter than 9th and brighter than 11th magnitude.
This is the transition area between positions based on
short exposures only and the combination of short and long exposures.
Also, in this magnitude range the combined formal error 
is lowest, allowing a moderate additional contribution from 
systematic errors to show up most pronounced.

At the faint end (V$\ge$11.0) the excess of the observed scatter
w.r.t.~the formal errors is relatively small; insignificant to
$\approx$10\% when compared with the $\sigma_{cx}, \sigma_{cy}$
(cut distribution), and about 20\% when compared to the
$\sigma_{qx}, \sigma_{qy}$ numbers.
This gives an external verification of the accuracy of the Tycho--2
astrometry and sets an upper limit to the ratio of true/formal
errors to about 1.2 at the faint end of the catalog, which
includes the majority of the stars.
At those magnitudes the UCAC1 positions, even if underestimated
by their formal errors, are more accurate than the Tycho--2 positions. 
The high UCAC1 accuracy is also confirmed by a comparison 
with SPM data \cite{ucac1}.

\section{Conclusions}

The Tycho--2 and UCAC1 positions overall are in good agreement with
systematic differences being very small by current astrometric catalog
\mbox{standards}. 
This external comparison shows the observed scatter in the
position differences to be 10 to 50\% larger
than the formal, internal precisions predict.
A significant fraction of the additional errors can be explained
by systematic errors present in the current UCAC1, which by 
all means is a preliminary catalog.
The UCAC1 contains the best positions available today
(at current epochs) for stars in the 11 to 16 magnitude range.
UCAC today covers only 80\% of the Southern Hemisphere with
no data in the North.
When completed in 2003 it will form the basis for the FAME \cite{fame}
input catalog.
The Tycho--2 Catalogue covers all sky and is the preferred
astrometric reference down to about magnitude V=10.5 or
wherever no UCAC1 data are available.



\acknowledgments

We wish to thank the entire UCAC project team and 
the CTIO staff, in particular our observers D.Castillo and M.Martinez.
More information on the UCAC project is available at
\url{http://ad.usno.navy.mil/ad/ucac/}.
The \mbox{UCAC1} is available on CD--ROM on request to
\email{nz@pisces.usno.navy.mil}.
The Tycho--2 Catalogue is available on CD--ROM on request to
\email{Tycho-2@astro.ku.dk} or 
\email{seu@pyxis.usno.navy.mil}.

\begin{figure}
\figurenum{1}
\epsscale{1.15}
\plotone{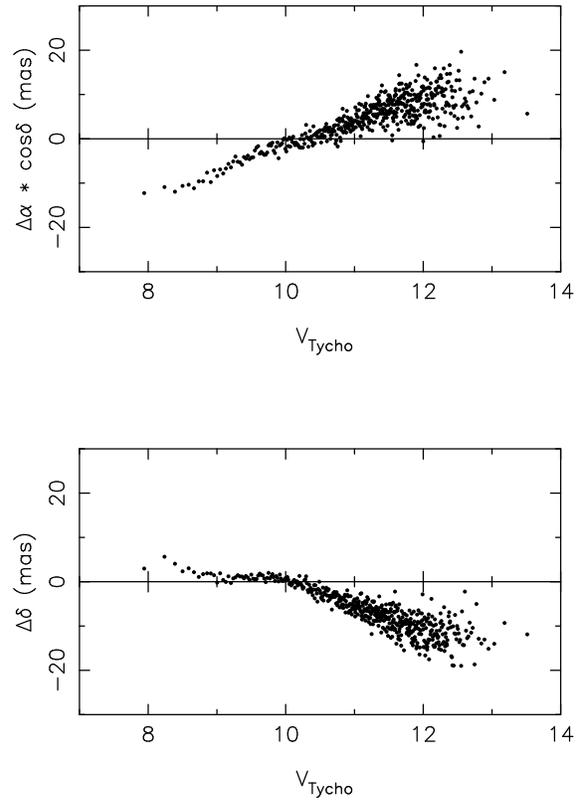}
\caption{Position differences (Tycho2--UCAC1)
  as a function of Tycho V magnitude.  All stars in common,
  which are not used as reference stars in the UCAC reduction
  are included.  One dot represents the mean for 2000 stars.}
\end{figure}

\begin{figure}
\figurenum{2}
\epsscale{1.15}
\plotone{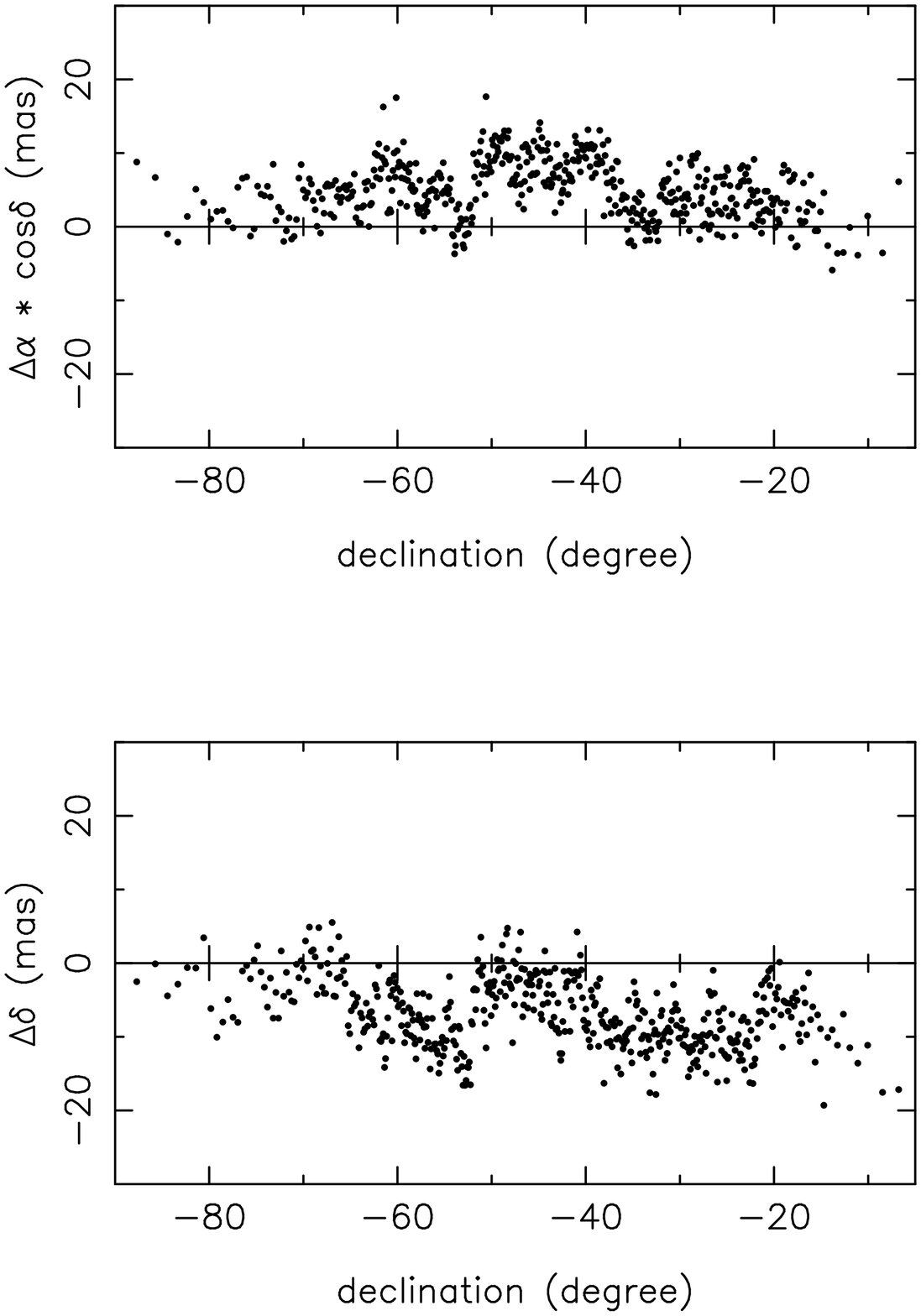}
\caption{Position differences (Tycho2--UCAC1)
  as a function of declination.
  One dot represents the mean for 2000 stars.}
\end{figure}

\begin{figure}
\figurenum{3}
\epsscale{1.15}
\plotone{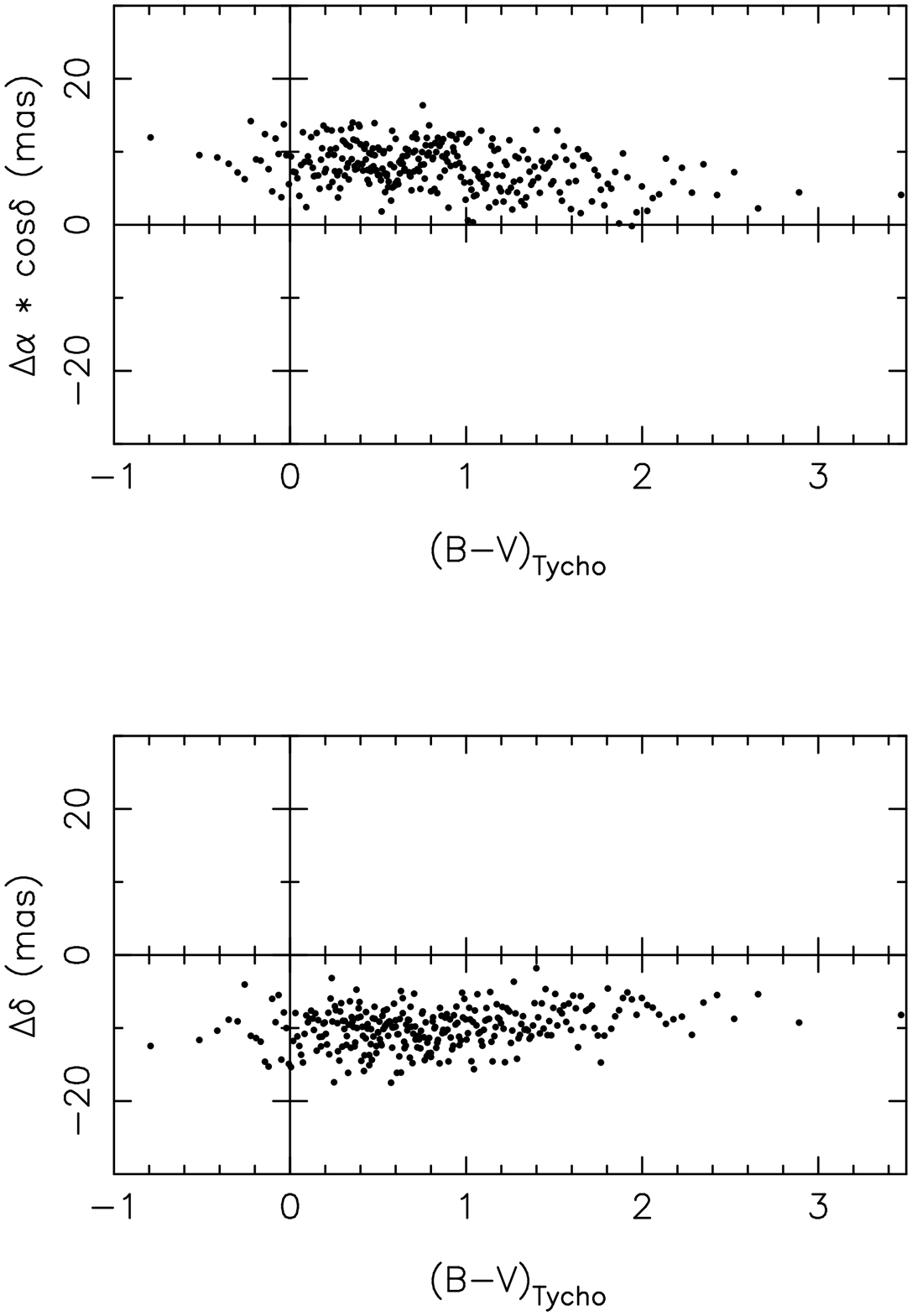}
\caption{Position differences (Tycho2--UCAC1)
  as a function of color index.
  One dot represents the mean of 2000 stars.}
\end{figure}




\begin{table}
\begin{center}
\caption{Internal precision of Tycho--2 ($\sigma_{T}$) and 
   UCAC1 ($\sigma_{U}$) positions as a function of Tycho visual ($V_{T}$)
   and UCAC red ($R_{U}$) magnitudes.
   The error contribution from the proper motions is given in column
   $\sigma_{PM}$, and $\sigma_{T-U}$ gives the formal expected error
   in the Tycho $-$ UCAC1 position comparison.}
\begin{tabular}{rrrrrr}
\tableline\tableline
$V_{T}$ & $\sigma_{T}$ & $\sigma_{PM}$ & $R_{U}$ & $\sigma_{U}$ & $\sigma_{T-U}$ \\
mag &  mas & mas & mag & mas & mas \\
\tableline
 8.0 &   5  &  10 &  7.5 & $\ge$60 & $\ge$61 \\
 8.5 &   6  &  10 &  8.0 &  40 &  42 \\
 9.0 &   9  &  11 &  8.5 &  35 &  38 \\
 9.5 &  14  &  12 &  9.0 &  30 &  35 \\
10.0 &  21  &  13 &  9.5 &  24 &  35 \\
10.5 &  32  &  15 & 10.0 &  16 &  39 \\
11.0 &  48  &  17 & 10.5 &  16 &  53 \\
11.5 &  70  &  19 & 11.0 &  16 &  74 \\
12.0 &  93  &  21 & 11.5 &  17 &  97 \\
12.5 & 110  &  24 & 12.0 &  17 & 114 \\
\tableline
\end{tabular}
\end{center}
\end{table}

\begin{table}
\begin{center}
\caption{Results of the catalog comparison using only
  the Tycho--2 stars not contained in Tycho--1 as a
  function of $V_{T}$ magnitude.  The column $n$ gives
  the number of stars in a given magnitude bin.
  The standard errors for $x$ and $y$ are for
  $\Delta\alpha \cos \delta$ and $\Delta\delta$
  respectively.
  For more details see the section 3.2.}
\begin{tabular}{rrrrrrr}
\tableline\tableline
 $V_{T}$ & n & $\sigma_{cx}$ & $\sigma_{cy}$ &
               $\sigma_{qx}$ & $\sigma_{qy}$ & $\sigma_{T-U}$ \\
 mag &   & mas & mas & mas & mas & mas \\
\tableline
  8.0 &      8 &  98 &  86 &  88 &  23 & $\ge$61 \\
  8.5 &     19 &  42 &  31 &  56 &  35 &  42 \\
  9.0 &     25 &  24 &  36 &  27 &  34 &  38 \\
  9.5 &     36 &  56 &  45 &  56 &  39 &  35 \\
 10.0 &    155 &  47 &  43 &  50 &  52 &  35 \\
 10.5 &   4088 &  59 &  49 &  72 &  61 &  39 \\
 11.0 &  58564 &  65 &  57 &  79 &  71 &  53 \\
 11.5 & 204984 &  79 &  72 &  96 &  88 &  74 \\
 12.0 & 209416 &  99 &  92 & 119 & 113 &  97 \\
 12.5 & 112272 & 118 & 111 & 143 & 137 & 114 \\
\tableline
\end{tabular}
\end{center}
\end{table}

\end{document}